\documentclass[prd,twocolumn,showpacs,amsmath,amssymb,letterpaper]{revtex4}
\usepackage{graphicx}
\usepackage{bm}
\begin{document}

\title{Supernova Neutrinos and the LSND Evidence for Neutrino Oscillations}

\author{Michel Sorel}
\email{sorel@fnal.gov}
\author{Janet Conrad}
\email{conrad@fnal.gov}
\affiliation{Department of Physics, Columbia University, New York, NY 10027}
\date{\today}

\begin{abstract}
The observation of the $\bar{\nu}_e$ energy spectrum from a
 supernova burst can provide constraints on neutrino oscillations.
 We derive formulas for adiabatic oscillations of supernova antineutrinos
 for a variety of 3 and 4-neutrino mixing schemes and
 mass hierarchies which are consistent with the LSND evidence for
 $\bar{\nu}_{\mu}\rightarrow \bar{\nu}_e$ oscillations. Finally, we
 explore the constraints on these models and LSND given by the
 supernova SN1987A $\bar{\nu}_e$'s observed by the Kamiokande-2 and
 IMB-3 detectors. 
\end{abstract}


\maketitle
\section{\label{sec:one}INTRODUCTION}
In recent years, the treatment of neutrino transport in the environment
 of a core-collapse supernova (SN) explosion
 has improved to the point of making realistic
 predictions on the observables for neutrinos reaching the Earth
 \cite{burrows,janka1,janka2,raffelt1}.
 Of particular interest for this paper are the average energies
 at the neutrinospheres, {\it i.e.} the surfaces
 of last scattering for the neutrinos, estimated to be
 $10-13$ MeV for $\nu_e$, $14-17$ MeV for $\bar{\nu}_e$, $23-27$ MeV
 for $\nu_{\mu ,\tau},\bar{\nu}_{\mu ,\tau}$ \cite{janka1,raffelt1}. \\
\indent The differences in temperatures between the various neutrino flavors
 can be qualitatively understood. Heavy-lepton neutrinos can interact only
 via neutral-current (NC) processes, the main contribution to their transport opacity
 coming from neutrino-nucleon scattering, which dominates over neutrino-electron
 scattering. In addition to this same NC contribution, the transport opacity for
 $\nu_e$'s and $\bar{\nu}_e$'s
 depends also on the charged-current (CC) absorptions $\nu_e+n\rightarrow
 p+e^-$ and $\bar{\nu}_e+p\rightarrow n+e^+$, respectively. Therefore, the
 $\nu_e$- and $\bar{\nu}_e$-spheres are located at larger radii with respect to
 the other neutrinospheres, that is at lower densities
 and lower temperatures. Moreover, in a neutron-rich environment,
 $\nu_e+ n\rightarrow p+e^-$ dominates over $\bar{\nu}_e+p\rightarrow
 n+e^+$: the emergent $\nu_e$'s originate from layers further
 outside the center of the star compared to $\bar{\nu}_e$'s, therefore at lower temperatures.
 The total energy released in a SN explosion is approximately equipartitioned
 between the different neutrino and antineutrino flavors \cite{janka2}. \\
\indent The above predictions can be confronted with the observation
 of the supernova $\bar{\nu}_e$ energy spectrum detected on Earth. Neutrino
 oscillations are expected to modify the spectrum since $\langle
 E_{\bar{\nu}_e} \rangle < \langle E_{\bar{\nu}_{\mu},\bar{\nu}_{\tau}}
 \rangle$. The energy dependence of the neutrino cross-section in the
 detector material, approximately $\sigma_{\bar{\nu}_e p}\propto
 (E_{\bar{\nu}_e}-1.29\mbox{MeV})^2$ \cite{beacom}, helps in making the
 $\bar{\nu}_e$ energy spectrum distortion a sensitive experimental
 probe to neutrino oscillations. This is because higher energy neutrinos interact
 significantly more than lower energy ones. \\
\indent We show that the extent of the spectrum modification depends
 crucially on the specifics of the neutrino mixing scheme and on the
 neutrino mass hierarchy under consideration, and we derive the
 relevant formulas assuming an adiabatic propagation for the antineutrinos
 in the supernova environment. Antineutrinos propagate adiabatically if the
 varying matter density they encounter changes slowly enough so that
 transitions between local (instantaneous) Hamiltonian eigenstates can
 be neglected throughout the entire antineutrino propagation. \\
\indent So far, neutrinos from one supernova have been detected and their
 energy measured:
 SN1987A was observed by the Kamiokande-2 and IMB-3 detectors. The overall
 20 events seen by those two detectors have all been
 interpreted as $\bar{\nu}_e$ interactions
 \cite{nuebar}.
 We examine the constraint of such observations
 on the LSND allowed region of $\bar{\nu}_{\mu}\rightarrow \bar{\nu}_e$
 oscillations \cite{lsnd}, for various neutrino mass and mixing models.
 If the LSND evidence is confirmed by
 the MiniBooNE experiment \cite{boone}, several models can be excluded
 or constrained on the basis of the observations of the
 supernova SN1987A and possibly future supernov\ae .
%
\section{\label{sec:two}ADIABATIC OSCILLATIONS AND NEUTRINO MIXING SCHEMES}
\subsection{\label{sec:p}$\bar{\nu}_e$ energy spectrum and the
 permutation factor}
In the presence of neutrino oscillations, the $\bar{\nu}_e$ flux
 reaching the Earth, $F_{\bar{\nu}_e}$, can be different from the primary
 flux at the neutrinosphere, $F^0_{\bar{\nu}_e}$. 
 We will assume that, at production, the
 energy of active antineutrinos is equally divided into the three active
 flavors, {\it i.e.} that
$\int_0^{\infty}\,dE_{\bar{\nu}_{\alpha}} E_{\bar{\nu}_{\alpha}} F^0_{\bar{\nu}_{\alpha}}$
 has the same numerical value for $\alpha =e,\mu ,\tau$.
 Moreover, we will also consider neutrino mixing models where the
 three
 active neutrino species are augmented by a fourth sterile neutrino with no
 standard weak couplings: in those cases, we will assume that the sterile
 component is negligible at production. \\
\indent The neutrino flux reaching the Earth is:
\begin{eqnarray}
F_{\bar{\nu}_e}=(p_{\mu\rightarrow e} + p_{\tau\rightarrow e})F^0_{\bar{\nu}_{\mu}}+
 p_{e\rightarrow e}F^0_{\bar{\nu}_e}
\nonumber \\
\propto (pF^0_{\bar{\nu}_{\mu}}+(1-p)F^0_{\bar{\nu}_e})
\label{eqn:flux}
\end{eqnarray}
\noindent where we have defined the {\it permutation factor} $p$ as:
\begin{equation}
p=\frac{p_{\mu \rightarrow e}+p_{\tau \rightarrow e}}{
 p_{\mu \rightarrow e}+p_{\tau \rightarrow e}+p_{e\rightarrow e}}
\label{eqn:swap}
\end{equation}
\noindent and $p_{\mu ,\tau , e\rightarrow e}$ are the probabilities
 for a $\bar{\nu}_{\mu}$, $\bar{\nu}_{\tau}$, $\bar{\nu}_e$ respectively at the
 neutrinosphere to oscillate into a $\bar{\nu}_e$.
 In Eqs.\ref{eqn:flux},\ref{eqn:swap}, 
 we have assumed that $p$ is energy-independent (as will be justified later), and that
 $\langle E_{\bar{\nu}_{\mu}} \rangle =
 \langle E_{\bar{\nu}_{\tau}}\rangle$.
In Eq.~\ref{eqn:flux}, we neglect
 the (energy-independent) proportionality factor since we will not deal with
 event rates, but only with neutrino energy distributions. 
%
%
\subsection{\label{sec:adiabatic}Neutrino propagation in the
 adiabatic approximation}
In vacuum, the Hamiltonian that governs neutrino propagation is diagonal
 in the mass eigenstate basis $|\nu_i \rangle$:
\begin{equation}
 (H_0)_{ij}\equiv \langle \nu_i|H_0|\nu_j \rangle = E_i \delta_{ij}
\end{equation}
\indent If the neutrinos all have the same relativistic momentum $p$,
 their energies
 $E_i$ differ only by a term proportional to their squared-mass differences,
 since $E_i\simeq p+m_i^2/2p$. If $U$ is the unitary mixing matrix that
 relates the
 flavor eigenstates $|\nu_{\alpha}\rangle$ to the mass eigenstates via
 $|\nu_{\alpha}\rangle = U_{\alpha i}|\nu_i \rangle$, the elements of the
 vacuum Hamiltonian in the flavor basis are given by
 \cite{kayser}:
\begin{equation}
(H_0)_{\alpha \beta}= U^*_{\alpha i}U_{\beta i}\frac{m_i^2}{2p}
\end{equation}    
\noindent where we have neglected the contribution $p\delta_{\alpha \beta}$
 in $(H_0)_{\alpha \beta}$, which is irrelevant for neutrino oscillations. \\
\indent In matter, $\bar{\nu}_e$'s undergo coherent CC forward-scattering
 from electrons, and all active flavor antineutrinos coherent NC
 forward-scattering from electrons, protons, and neutrons in the medium. These
 processes give rise to an interaction potential $V=V_W+V_Z$, which is
 diagonal in the flavor basis and proportional to the matter density
 $\rho$:
\begin{equation}
(V)_{\alpha \beta}=A_{\alpha} \frac{G_F\rho}{m_N} \delta_{\alpha \beta}
\end{equation}
\noindent where $A_{\alpha}$ is a proportionality constant, in general
 different for $\alpha = e, \mu , \tau, \mbox{or } s$, $G_F$ the
 Fermi constant, and $m_N$ the nucleon mass. The relevant
 Hamiltonian for neutrino propagation in matter is therefore $H\equiv
 H_0+V$. \\
\indent At the neutrinosphere, the density $\rho$ is so high ($\sim 10^{12}\
 g/cm^3$ \cite{burrows}) that the interaction potential dominates over the
 vacuum Hamiltonian, so that the propagation eigenstates coincide with the
 flavor eigenstates. As the propagation eigenstates free-stream outwards,
 toward regions of lower density, their flavor composition changes,
 ultimately reaching the flavor composition of the mass eigenstates in the
 vacuum. Given that the neutrinos escape the SN as mass eigenstates, no
 further flavor oscillations occur on their path to the Earth. \\
\indent More specifically, making use of the adiabatic approximation and of
 the fact that no
 energy-level crossing is permitted, the flavor eigenstate at the
 neutrinosphere with the
 maximum interaction potential reaches Earth as the mass eigenstate with the
 biggest neutrino mass. In general, the energy level order is maintained
 throughout the neutrino propagation in the SN ejecta. \\
\indent This is illustrated
 in Tab.~\ref{tab:adiabatic} for three neutrinos in the row labelled
 ``Normal $(1+1+1)$'', where we have taken
 $A_{\gamma}>A_{\beta}>A_{\alpha}$ and $m_3>m_2>m_1$.
\begin{table}[bt]
\begin{ruledtabular}
\begin{tabular}{lcr}
Model & Hierarchy & Propagation \\ \hline
Normal (1+1+1)& $m_3>m_2>m_1$ & $\bar{\nu}_{\gamma}\rightarrow \bar{\nu}_3$ \\
 & & $\bar{\nu}_{\beta}\rightarrow \bar{\nu}_2$ \\
 & & $\bar{\nu}_{\alpha}\rightarrow \bar{\nu}_1$ \\ \hline \hline
Normal (1+1)& $m_2\gg m_1$ & $\bar{\nu}_{\mu}\rightarrow \bar{\nu}_2$ \\
 & & $\bar{\nu}_{e}\rightarrow \bar{\nu}_1$ \\ \hline
LSND-inverted (1+1) & $m_1\gg m_2$ & $\bar{\nu}_{\mu}\rightarrow \bar{\nu}_1$ \\
 & & $\bar{\nu}_{e}\rightarrow \bar{\nu}_2$ \\ \hline
Normal (2+1) & $m_3>m_2\gg m_1$ & $\bar{\nu}_{\mu}\rightarrow \bar{\nu}_3$ \\
 & & $\bar{\nu}_{\tau}\rightarrow \bar{\nu}_2$ \\
 & & $\bar{\nu}_{e}\rightarrow \bar{\nu}_1$ \\ \hline
LSND-inverted (2+1) & $m_1\gg m_3>m_2$ & $\bar{\nu}_{\mu}\rightarrow \bar{\nu}_1$ \\
 & & $\bar{\nu}_{\tau}\rightarrow \bar{\nu}_3$ \\
 & & $\bar{\nu}_{e}\rightarrow \bar{\nu}_2$ \\ \hline
Normal (2+2)& $m_3>m_2\gg m_1>m_0$ & $\bar{\nu}_{\mu}\rightarrow \bar{\nu}_3$ \\
 & & $\bar{\nu}_{\tau}\rightarrow \bar{\nu}_2$ \\
 & & $\bar{\nu}_{s}\rightarrow \bar{\nu}_1$ \\
 & & $\bar{\nu}_{e}\rightarrow \bar{\nu}_0$ \\ \hline
LSND-inverted (2+2) & $m_1>m_0\gg m_3>m_2$ & $\bar{\nu}_{\mu}\rightarrow
 \bar{\nu}_1$ \\
 & & $\bar{\nu}_{\tau}\rightarrow \bar{\nu}_0$ \\
 & & $\bar{\nu}_{s}\rightarrow \bar{\nu}_3$ \\
 & & $\bar{\nu}_{e}\rightarrow \bar{\nu}_2$ \\ \hline
Normal (3+1) & $m_4\gg m_3>m_2>m_1$ & $\bar{\nu}_{\mu}\rightarrow \bar{\nu}_4$ \\
 & & $\bar{\nu}_{\tau}\rightarrow \bar{\nu}_3$ \\
 & & $\bar{\nu}_{s}\rightarrow \bar{\nu}_2$ \\
 & & $\bar{\nu}_{e}\rightarrow \bar{\nu}_1$ \\ \hline
LSND-inverted (3+1) & $m_3>m_2>m_1\gg m_4$ & $\bar{\nu}_{\mu}\rightarrow
 \bar{\nu}_3$ \\
 & & $\bar{\nu}_{\tau}\rightarrow \bar{\nu}_2$ \\
 & & $\bar{\nu}_{s}\rightarrow \bar{\nu}_1$ \\
 & & $\bar{\nu}_{e}\rightarrow \bar{\nu}_4$ \\ 

\end{tabular}
\end{ruledtabular}
\caption{\label{tab:adiabatic}Adiabatic neutrino propagation in the SN ejecta
 for the neutrino mixing models considered.}
\end{table}
 For example, the probability for a $\bar{\nu}_{\alpha}$ to emerge from
 the SN environment as a $\bar{\nu}_{\beta}$ is given by:
\begin{eqnarray}
p_{\alpha \rightarrow \beta} =
 |\langle \bar{\nu}_{\beta}|U^{evol}|\bar{\nu}_{\alpha}\rangle |^2 =
 | \langle U_{\beta i}\bar{\nu}_i|U^{evol}|\bar{\nu}_{\alpha}\rangle |^2 =
 \nonumber \\
 |U_{\beta i}^* \delta_{i,1}|^2 = |U_{\beta 1}|^2
\label{eqn:palphabeta}
\end{eqnarray}
\noindent where $U^{evol}$ is the adiabatic evolution operator. In
 Eq.\ref{eqn:palphabeta}, we have used Tab.~\ref{tab:adiabatic} to get:
\begin{equation}
\langle \bar{\nu}_i|U^{evol}|\bar{\nu}_{\alpha}\rangle = \delta_{i,1}
\end{equation}
\indent This result can be immediately generalized to any number of
 antineutrino generations.
 Also, as long as the adiabatic approximation is satisfied, the formula does
 not depend on the specific dynamics for the neutrino propagation, for
 example on the number and position in the SN environment of
 MSW-resonances. We will comment more on the validity of the adiabatic
 approximation in the next section. \\
\indent In this paper, we consider three or four flavor components,
 including a sterile one. At tree-level, the proportionality factors
 $A_{\alpha}$ in the interaction potential for neutral matter are
 \cite{kayser,caldwell}:
\begin{equation}
A=
\left\{
\begin{array}{ll}
(1-3Y_e)/\sqrt{2}, & \mbox{for }\bar{\nu}_e \\
(1-Y_e)/\sqrt{2}, & \mbox{for }\bar{\nu}_{\mu},\bar{\nu}_{\tau} \\
0, & \mbox{for }\bar{\nu}_s \\
\end{array}
\right.
\label{eqn:a}
\end{equation}   
\noindent where $Y_e$ is the electron fraction per nucleon.
 Following the assumptions of
 \cite{caldwell, qian}, we use $Y_e \simeq (1+\langle E_{\bar{\nu}_e}\rangle
 /\langle E_{\nu_e}\rangle)^{-1} > 1/3$ at the neutrinosphere.
 Considering also one-loop electroweak radiative corrections, a
 difference in the $\bar{\nu}_{\mu}$ and $\bar{\nu}_{\tau}$ interaction
 potentials of magnitude $(A_{\mu}-A_{\tau})/A_{\mu}\sim 10^{-4}$
 appears due to the difference in the charged lepton masses
 \cite{botella, dighe}. At the neutrinosphere,
 this second-order effect in the interaction potential dominates over
 the vacuum Hamiltonian terms (as long as $|m_i^2-m_j^2|<10\ eV^2$ for all
 $i,j$), and removes the
 $\bar{\nu}_{\mu}-\bar{\nu}_{\tau}$ degeneracy. Therefore, for the
 antineutrino channel considered here, we take:
\begin{equation}
 A_{\mu}>A_{\tau}>A_s>A_e
\label{eqn:order}
\end{equation}
\noindent For the neutrino channel, one should substitute $A\rightarrow -A$
 in Eq.\ref{eqn:a}, and the order in Eq.\ref{eqn:order} would be inverted. \\
\indent Therefore, given a specific neutrino mass and mixing model,
 the permutation factor can be easily evaluated in the adiabatic
 approximation, and its numerical value does not depend on the neutrino
 energy. We will comment on possible energy-dependent Earth matter
 effects in the next section. In practice, one proceeds backwards: given
 a certain measured value of $p$, it is possible to constrain possible
 models for neutrino oscillations. This approach is used for example in \cite{dighe}
 to constrain models explaining the solar and atmospheric neutrino data;
 in this paper, we focus on 3 and 4-neutrino models explaining the LSND
 data.
%
%
\subsection{\label{sec:mixing}Possible mixing schemes}
 The results for the $\bar{\nu}_{\mu}$,$\bar{\nu}_{\tau}$,$\bar{\nu}_e
 \rightarrow \bar{\nu}_e$ adiabatic oscillation probabilities, the
 permutation factor $p$, and the LSND oscillation amplitude
 $\sin^2 2\vartheta$ as a function of the mixing parameters and $p$ for
 the eight possible mass and mixing schemes considered below are given in
 Tab.~\ref{tab:results}. The mass hierarchy and the adiabatic propagation of
 the neutrino eigenstates for these mixing schemes are depicted in
 Tab.~\ref{tab:adiabatic}. \\
%
%
\begin{table*}[bt]
\begin{ruledtabular}
\begin{tabular}{lllllll}
Model & Mixing & $p_{\mu \rightarrow e}$ & $p_{\tau \rightarrow e}$ &
 $p_{e \rightarrow e}$ & $p$ & $\sin^2 2\vartheta_{LSND}$ \\ \hline
Normal (1+1) & Eq.\ref{eqn:1+1} & $\sin^2\vartheta$ & $0$ & $\cos^2\vartheta$
 & $\sin^2 \vartheta$ & $\sin^2 2\vartheta = 4p(1-p)$ \\
LSND-inverted (1+1) & Eq.\ref{eqn:1+1} & $\cos^2\vartheta$ & $0$ & $\sin^2\vartheta$
 & $\cos^2 \vartheta$ & $\sin^2 2\vartheta = 4p(1-p)$ \\
Normal (2+1) & Eq.\ref{eqn:2+1} & $\frac{3}{4}\alpha^2$ & $\frac{1}{4}\alpha^2$ &
 $1$
 & $\alpha^2/(1+\alpha^2)$ &
 $4\alpha^2 = 4p/(1-p)$ \\
LSND-inverted (2+1) & Eq.\ref{eqn:2+1} & $1$ & $\frac{3}{4}\alpha^2$ &
 $\frac{1}{4}\alpha^2$ & $(1+\frac{3}{4}\alpha^2)/(1+\alpha^2)$ &
 $4\alpha^2 = 4(1-p)/(p-\frac{3}{4})$ \\
Normal (2+2) & Eq.\ref{eqn:2+2mod} & $\beta^2$ & $\beta^2$ & $\frac{1}{2}$
 & $4\beta^2/(1+4\beta^2)$ &
 $8\beta^2 = 2p/(1-p)$ \\
LSND-inverted (2+2) & Eq.\ref{eqn:2+2mod} & $\frac{1}{2}$ & $\frac{1}{2}$ &
 $\beta^2$ & $1/(1+\beta^2)$ &
 $8\beta^2 = 8(1-p)/p$ \\
Normal (3+1) & Eq.\ref{eqn:3+1} & $\gamma^2$ & $0$ & $\frac{1}{2}$
 & $2\gamma^2/(1+2\gamma^2)$ &
 $4\gamma^2\delta^2 = 2\delta^2 p/(1-p)$ \\
LSND-inverted (3+1) & Eq.\ref{eqn:3+1} & $0$ & $\frac{1}{2}$ & $\gamma^2$
 & $1/(1+2\gamma^2)$ &
 $4\gamma^2\delta^2 = 2\delta^2 (1-p)/p$ \\
\end{tabular}
\end{ruledtabular}
\caption{\label{tab:results}Results on the probabilities
 $p_{\mu , \tau , e\rightarrow
 e}$ for a $\bar{\nu}_{\mu , \tau , e}$ to emerge from the SN as a
 $\bar{\nu}_e$, the permutation factor $p$ of Eq.\ref{eqn:swap}, and
 the LSND oscillation amplitude $\sin^2 2\vartheta_{LSND}$, for
 the various neutrino mixing schemes considered.}
\end{table*}
%
%
\indent The simplest possible mixing scheme is a $(1+1)$ model explaining only
 $\bar{\nu}_{\mu} \rightarrow \bar{\nu}_e$ LSND oscillations in vacuum,
 and not the atmospheric or solar oscillations:
\begin{equation}
\left(
\begin{array}{l}
\bar{\nu}_e \\
\bar{\nu}_{\mu} \\
\end{array}
\right)
=
\left(
\begin{array}{rr}
\cos\vartheta & \sin\vartheta \\
-\sin\vartheta & \cos\vartheta \\
\end{array}
\right)
\left(
\begin{array}{l}
\bar{\nu}_1 \\
\bar{\nu}_2 \\
\end{array}
\right)
\label{eqn:1+1}
\end{equation}
\noindent where the mixing angle $\vartheta$ can assume any value in the range
 $0<\vartheta <\pi /4$. \\
\indent We consider a $(2+1)$ model motivated, for example, by CPT-violating scenarios
 (see, e.g. \cite{barenboim1}), in which atmospheric and LSND oscillations
 in the antineutrino channel are obtained via the mixing
 \cite{barenboim2}:
\begin{equation}
\left(
\begin{array}{l}
\bar{\nu}_{e} \\
\bar{\nu}_{\mu} \\
\bar{\nu}_{\tau} \\
\end{array}
\right)
=
\left(
\begin{array}{ccc}
1 & -\frac{1}{2}\alpha & -\frac{\sqrt{3}}{2}\alpha \\
\alpha & \frac{1}{2} & \frac{\sqrt{3}}{2} \\
0 & -\frac{\sqrt{3}}{2} & \frac{1}{2} \\
\end{array}
\right)
\left(
\begin{array}{l}
\bar{\nu}_1 \\
\bar{\nu}_2 \\
\bar{\nu}_3
\end{array}
\right)
\label{eqn:2+1}
\end{equation}
\indent The matrix in Eq.\ref{eqn:2+1} is chosen to ensure large $\bar{\nu}_{\mu}
 \rightarrow \bar{\nu}_{\tau}$ mixing for atmospheric neutrinos
 ($\sin^2 2\vartheta_{atm}=3/4$), while the LSND $\bar{\nu}_{\mu}\rightarrow \bar{\nu}_e$
 mixing is fixed by the parameter $\alpha$ ($\sin^2 2\vartheta_{LSND}=
 4\alpha^2$). \\
\indent The most popular models which explain the solar, atmospheric and LSND
 signatures (and the null results obtained by other experiments)
 via neutrino oscillations invoke the existence of a sterile
 neutrino $\bar{\nu}_s$. One example of a $(2+2)$ model is
 the following, which is taken from \cite{barger}:
\begin{equation}
\left(
\begin{array}{l}
\bar{\nu}_{s} \\
\bar{\nu}_{e} \\
\bar{\nu}_{\mu} \\
\bar{\nu}_{\tau} \\
\end{array}
\right)
=
\left(
\begin{array}{cccc}
\frac{1}{\sqrt{2}} & \frac{1}{\sqrt{2}} & 0 & 0 \\
-\frac{1}{\sqrt{2}} & \frac{1}{\sqrt{2}} & \beta & \beta \\
\beta & -\beta & \frac{1}{\sqrt{2}} & \frac{1}{\sqrt{2}} \\
0 & 0 & -\frac{1}{\sqrt{2}} & \frac{1}{\sqrt{2}} \\ 
\end{array}
\right)
\left(
\begin{array}{l}
\bar{\nu}_0 \\
\bar{\nu}_1 \\
\bar{\nu}_2 \\
\bar{\nu}_3 \\
\end{array}
\right)
\label{eqn:2+2}
\end{equation} 
\noindent  where one pair of nearly degenerate mass eigenstates has maximal
 $\nu_e \rightarrow \nu_s$ mixing for solar neutrinos and the other pair
 has maximal $\nu_{\mu}\rightarrow \nu_{\tau}$ mixing for atmospheric
 neutrinos. Small inter-doublet mixings through the
 $\beta$ parameter accomodates the
 LSND result ($\sin^2 \vartheta_{LSND}=8\beta^2$). \\
\indent Recent experimental results \cite{sno} show that pure
 $\nu_e\rightarrow \nu_s$ solar oscillations are excluded at high
 significance. We therefore consider a more general $(2+2)$ scenario,
 in which solar neutrinos can undergo any combination of
 $\nu_e\rightarrow \nu_s$ and $\nu_e\rightarrow \nu_{\tau}$ oscillations,
 while atmospheric neutrinos can undergo any combination of
 $\nu_{\mu}\rightarrow \nu_{\tau}$ and $\nu_{\mu}\rightarrow \nu_s$
 oscillations. We follow the procedure in \cite{barger2} to obtain this
 more general mixing starting from Eq.\ref{eqn:2+2}, by substituting
 the $(\bar{\nu}_s,\bar{\nu}_{\tau})$ states with the rotated states
 $(\bar{\nu}'_s,\bar{\nu}'_{\tau})$:
\begin{equation}
\left(
\begin{array}{l}
\bar{\nu}'_s \\
\bar{\nu}'_{\tau} \\
\end{array}
\right)
=
\left(
\begin{array}{rr}
\cos\varphi_s & \sin\varphi_s \\
-\sin\varphi_s & \cos\varphi_s \\
\end{array}
\right)
\left(
\begin{array}{l}
\bar{\nu}_s \\
\bar{\nu}_{\tau} \\
\end{array}
\right)
\label{eqn:steriletau}
\end{equation} 
\noindent where the rotation angle $\varphi_s$ fixes the sterile
 component in the atmospheric doublet ($0<\varphi_s<\pi /2$).
 Eq.\ref{eqn:2+2} then becomes:
\begin{equation}
\left(
\begin{array}{l}
\bar{\nu}_{s} \\
\bar{\nu}_{e} \\
\bar{\nu}_{\mu} \\
\bar{\nu}_{\tau} \\
\end{array}
\right)
=
\left(
\begin{array}{cccc}
\frac{\cos\varphi_s}{\sqrt{2}} & \frac{\cos\varphi_s}{\sqrt{2}} &
 \frac{\sin\varphi_s}{\sqrt{2}} & -\frac{\sin\varphi_s}{\sqrt{2}}  \\
-\frac{1}{\sqrt{2}} & \frac{1}{\sqrt{2}} & \beta & \beta \\
\beta & -\beta & \frac{1}{\sqrt{2}} & \frac{1}{\sqrt{2}} \\
\frac{\sin\varphi_s}{\sqrt{2}} & \frac{\sin\varphi_s}{\sqrt{2}} &
 -\frac{\cos\varphi_s}{\sqrt{2}} & \frac{\cos\varphi_s}{\sqrt{2}} \\ 
\end{array}
\right)
\left(
\begin{array}{l}
\bar{\nu}_0 \\
\bar{\nu}_1 \\
\bar{\nu}_2 \\
\bar{\nu}_3 \\
\end{array}
\right)
\label{eqn:2+2mod}
\end{equation} 
\noindent which contains Eq.\ref{eqn:2+2} in the specific case
 $\varphi_s=0$. We note that the LSND oscillation amplitude formula
 $\sin^2 2\vartheta_{LSND}=8\beta^2$ holds also for the more general
 case of Eq.\ref{eqn:2+2mod}, and that our results are independent
 of the value of $\varphi_s$ (see Tab.~\ref{tab:results}). \\
\indent Another possible 4-neutrino model has a $(3+1)$ hierarchy; as an
 example for this model, here we
 consider the following mixing, which is also taken from
 \cite{barger}:
\begin{equation}
\left(
\begin{array}{l}
\bar{\nu}_{e} \\
\bar{\nu}_{\mu} \\
\bar{\nu}_{\tau} \\
\bar{\nu}_{s} \\
\end{array}
\right)
=
\left(
\begin{array}{cccc}
\frac{1}{\sqrt{2}} & \frac{1}{\sqrt{2}} & 0 & \gamma \\
-\frac{1}{2} & \frac{1}{2} & \frac{1}{\sqrt{2}} & \delta \\
\frac{1}{2} & -\frac{1}{2} & \frac{1}{\sqrt{2}} & 0 \\
\frac{1}{2}\delta -\frac{1}{\sqrt{2}}\gamma &
 -\frac{1}{2}\delta -\frac{1}{\sqrt{2}}\gamma & -\frac{1}{\sqrt{2}}\delta &
 1 \\ 
\end{array}
\right)
\left(
\begin{array}{l}
\bar{\nu}_1 \\
\bar{\nu}_2 \\
\bar{\nu}_3 \\
\bar{\nu}_4 \\
\end{array}
\right)
\label{eqn:3+1}
\end{equation} 
\noindent where the solar and atmospheric oscillations are approximately
 described by oscillations of three active neutrinos, and the LSND result
 by a coupling of $\bar{\nu}_{\mu}$ and $\bar{\nu}_e$ through
 small mixings with $\bar{\nu}_s$ that has a mass eigenvalue widely
 separated from the others ($\sin^2 2\vartheta_{LSND}=4\gamma^2\delta^2$).
 For the $(3+1)$ scenario, the constraint given by the permutation probability
 $p$ is not sufficient to determine the LSND oscillation amplitude
 $\sin^2 2\vartheta_{LSND}$. Therefore, the constraint on
 $|U_{\mu 4}|^2=\delta^2$ given by
 the CDHS and Super-K experiments will also be used, as explained
 later. \\
\indent We should note that the mixing matrices defined in Eqs.~\ref{eqn:1+1}-
\ref{eqn:3+1} are approximations in the sense that the matrices are unitary
 only up to order $\mathcal{O}(\alpha ,\beta ,\gamma ,\delta )$. These are
 the parameters in the mixings responsible for LSND-type oscillations, which we
 let float for our analysis, but we know they are small. \\
\begin{figure}[tb]
\includegraphics{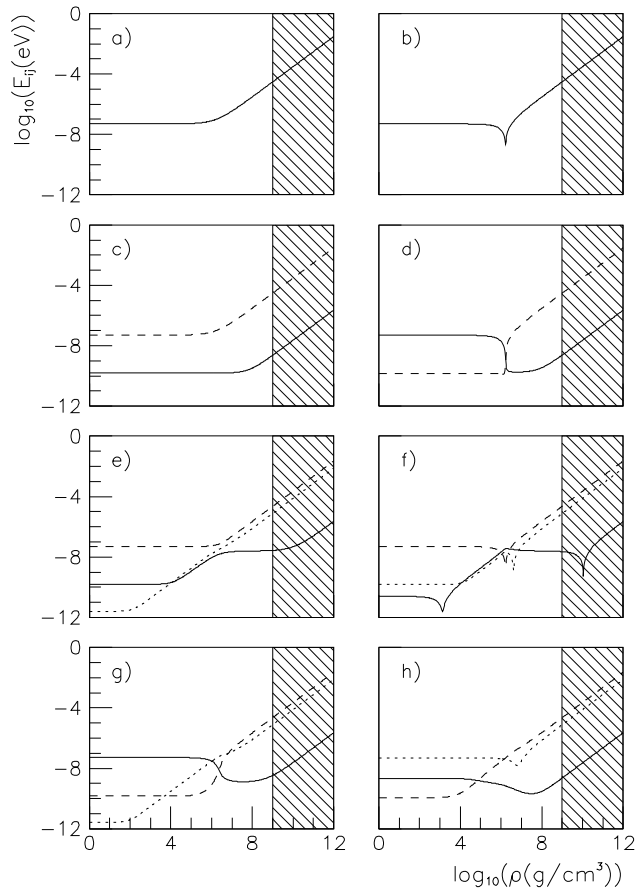}
\caption{\label{fig:respos}Splittings between energy eigenvalues versus matter
 density $\rho$ for various neutrino mass and mixing models.
 Solid, dashed, dotted lines show the splittings $E_{12}$,
 $E_{23}$, $E_{34}$, respectively (see text). The local minima
 correspond to MSW-resonances. Model: a)normal (1+1); b)inverted (1+1);
 c)normal (2+1); d)inverted (2+1); e)normal (2+2); f)inverted (2+2); g)normal (3+1);
 h)inverted (3+1). Apart from the inconsequential $\bar{\nu}_{\mu}\leftrightarrow
 \bar{\nu}_{\tau}$ one in Fig.\ref{fig:respos}f, no MSW-resonances occur before the
 antineutrinos reach the stalled shock-wave (hatched area).}
\end{figure} 
\indent In order to determine the permutation factor for the mixing models,
 we also need to specify the neutrino mass hierarchy. In this paper, we
 consider for each mixing model both the cases of a ``normal'' and a
 ``LSND-inverted'' mass hierarchies. By ``normal'' hierarchy, here we mean that
 $m_i>m_j$ for $i>j$, where $m_i$ is the mass eigenvalue for the
 $|\bar{\nu}_i\rangle$ state. We define the ``LSND-inverted'' hierarchies
 as the ones obtained substituting $\Delta m_{LSND}\rightarrow
 -\Delta m_{LSND}$ in the normal hierarchies, without changing the hierarchy of
 the eventual solar and atmospheric splittings (see Tab.~\ref{tab:adiabatic});
 $\Delta m_{LSND}$ is the neutrino mass difference responsible for
 LSND oscillations. \\  
\indent A common feature to all the mixing schemes is apparent in
 Tab.~\ref{tab:results}. In the adiabatic
 approximation, normal mass hierarchies predict small permutation factors,
 while an almost complete permutation would be present for
 LSND-inverted hierarchies. \\
\indent Given the specific neutrino mixing models considered here, we can
 now partially address the question whether the
 adiabatic approximation is applicable in this context.
 At a resonance, where the non-adiabaticity is maximal, this is a
 good approximation if
 the width of the resonance region is large compared with the local
 neutrino oscillation length. The width of the resonance is, in turn,
 determined by the characteristic length scale of the radial matter
 density variations at the resonance. While there are reliable models for the
 matter density profile of the progenitor star, there are still uncertainties
 on the profile seen by neutrinos in their free-streaming propagation. \\
\indent
 It is now thought that neutrino heating of the proto-neutron star
 mantle drives the supernova explosion, which would happen with a $\sim 1s$
 delay after the creation of the shock-wave, ultimately  responsible for the
 explosion; during this delay, the shock-wave would be stalled
 at a radius of $\sim 200\ km$ from the neutron star, corresponding to
 a density $\rho\sim 10^9-10^{10}\ g/cm^3$ \cite{burrows}. Therefore, the
 density profile in the proximity of the stalled shock-wave, which is difficult
 to model reliably, is a potential site for non-adiabatic oscillations. \\
\indent In Fig.~\ref{fig:respos} we show the energy splittings between the
 local neutrino energy eigenvalues $E_i$, as a function of matter
 density, for all eight neutrino models considered here. For an
 $n$-neutrino model, we plot $E_{i,i+1}\equiv E_i-E_{i+1}$,
 where $i=1,\ldots ,n-1$; the eigenvalues are ordered such that
 $E_1>E_2>\ldots >E_n$. Clearly, a resonance corresponds to a local minimum
 in one of the curves. As can be seen from Fig.~\ref{fig:respos},
 all the resonances (except the inconsequential one in Fig.\ref{fig:respos}f
 between $\bar{\nu}_{\mu}$ and $\bar{\nu}_{\tau}$ \cite{2+2nonadiab})
 lie at densities well below the stalled shock-wave density of
 $\rho\sim 10^9-10^{10}\ g/cm^3$. Therefore, the impact of level
 crossing between propagation eigenstates is likely to be small even where
 the neutrinos encounter the shock-wave. \\  
\indent If the SN neutrinos cross the Earth on their way to the detector,
 as for example happened for the SN1987A $\bar{\nu}_e$'s detected by the
 Kamiokande-2 and IMB-3 detectors, it is also necessary to evaluate the importance
 of Earth matter effects in the neutrino propagation. Clearly, for neutrino
 oscillation models where no solar splitting is involved (for example
 the $(1+1)$ and $(2+1)$ models in this paper), this effect is negligible.
 In the models where such a splitting is allowed ({\it i.e.} the
 $(2+2)$ and $(3+1)$ models considered here), the situation is more
 complicated. However, the Earth matter effects have been shown to be
 small in this case as well for a large fraction of the SN $\bar{\nu}_e$
 energy spectrum (below $\simeq 40$ MeV)
 \cite{dighe, lunardini}, and for the sake of simplicity will not be considered
 further.
\begin{figure*}[tb]
\includegraphics{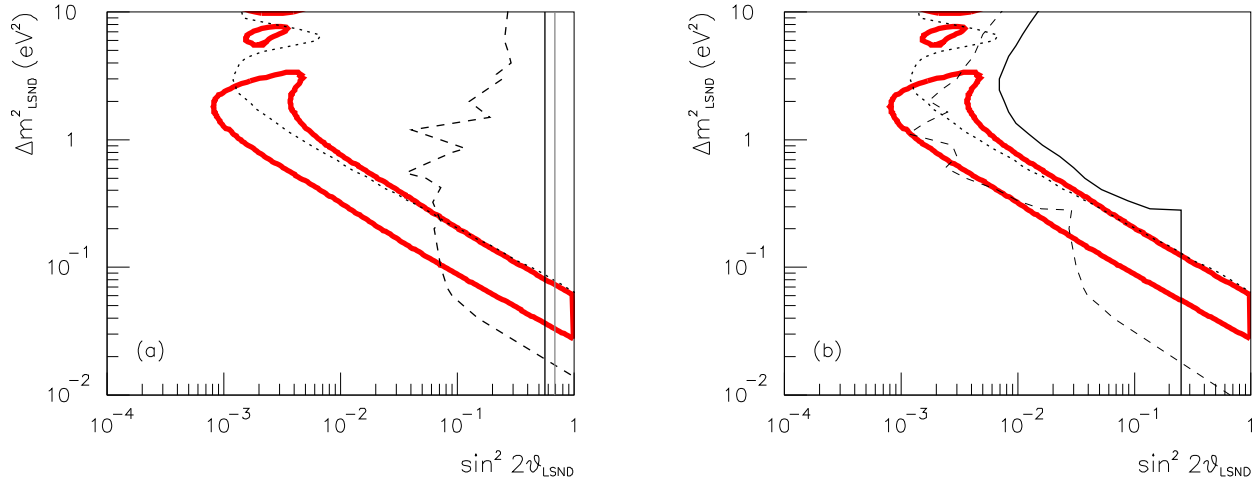}
\caption{\label{fig:parspace}$99\%$ CL LSND allowed region \cite{lsnd}
 and $99\%$ CL exclusion
 regions for the neutrino mixing schemes considered in the text and with
 normal mass hierarchy. The exclusion regions are estimated as in
 \cite{peres}. a) shows the exclusion regions for the $(1+1)$, $(2+1)$ and
 $(2+2)$ models, b) for the $(3+1)$ model. 
 The exclusion regions refer to experimental
 data from the following experiments. a) Dotted line: Karmen;
 dashed line: Bugey; dark solid line: SN1987A for the $(2+2)$ model;
 light solid line: SN1987 for the $(1+1)$ model;
 SN1987A data
 provide no constraints at $99\%$ CL for the $(2+1)$ model. b)
 Dotted line: Karmen; dashed line: Bugey, CDHS and Super-K;
 solid line: SN1987A, CDHS and Super-K.}
\end{figure*}
%
%
\section{\label{sec:three}CONSTRAINTS ON LSND FROM SN1987A OBSERVATIONS}
Twenty $\bar{\nu}_e$ events from the supernova SN1987A were observed by the
 Kamiokande-2 (Kam-2) and IMB-3 detectors. Kam-2 saw 12 events with an
 average energy of $\langle E_{det}\rangle = 14.7$ MeV, IMB-3 (which had
 a higher energy threshold than Kam-2) detected 8 events with
 $\langle E_{det}\rangle = 31.9$ MeV \cite{koshiba}. \\
\indent From a comparison of the measured energy spectra ($F_{\bar{\nu}_e}$)
 with theoretical models of neutrino emission ($F^0_{\bar{\nu}_{e}}$ and
 $F^0_{\bar{\nu}_{\mu}}$), it is possible to infer the permutation
 factor $p$ in Eq.~\ref{eqn:flux}. SN1987A observations are consistent
 with no-oscillations ({\it i.e.} $p=0$). In Appendix A, we derive
 a conservative upper bound on $p$ of $p<0.22$ at $99\%$ CL, by applying a
 Kholmogorov-Smirnov test on the joint Kam-IMB dataset and a range of
 supernova neutrino emission models. \\
\indent One important result of our analysis is immediately apparent
 from the
 values of the permutation factors $p$ as a function of the mixing
 parameters in Tab.~\ref{tab:results}, and from the fact that the value of $p$
 inferred from SN1987A data has to be less than $0.22$ at $99\%$ CL.
 The four mixing schemes considered,
 explaining the LSND effect via a LSND-inverted neutrino mass hierarchy,
 are all
 incompatible with SN1987A data. \\
\indent We now consider the normal hierarchy cases. For the $(1+1)$, $(2+1)$ and
 $(2+2)$
 models with the mixings of Eqs.~\ref{eqn:1+1}-\ref{eqn:2+2}, the bound on
 the permutation
 factor $p$ unambigously determines the constraint on the LSND oscillation
 amplitude $\sin^2 2\vartheta_{LSND}$ (see Tab.~\ref{tab:results}).
 At $99\%$ CL, SN1987A data provide no constraints on the $(2+1)$
 model, and a constraint which is weaker than existing bounds from the
 accelerator experiment Karmen \cite{karmen} and the reactor experiment Bugey
 \cite{bugey,bilenky} for the $(1+1)$ and $(2+2)$ models (see Fig.~\ref{fig:parspace}a).
 Therefore, these models are compatible with the SN1987A data. \\
\indent As already mentioned, for the $(3+1)$ model, the permutation factor
 does not fully determine
 the LSND oscillation amplitude: $\sin^2 2\vartheta_{LSND}$ depends not
 only on $p$, but also on $|U_{\mu 4}|^2=\delta^2$. Here we use the
 $\Delta m^2_{LSND}$-dependent constraints on $\delta^2$ from the
 $\nu_{\mu}$-disappearance experiments CDHS (for $\Delta m^2_{LSND}>0.3\ eV^2$)
 and Super-K (for $\Delta m^2_{LSND}<0.3\ eV^2$) \cite{bilenky}.
 Moreover, another
 complication arises in evaluating exclusion regions for (3+1) models:
 given the $99\%$ CL upper bounds on $\gamma^2 =|U_{e 4}|^2$ from
 SN1987A and $\delta^2 =|U_{\mu 4}^2|$ from CDHS and Super-K, what is
 the $99\%$ CL upper bound on $\sin^2 2\vartheta_{LSND}=4\gamma^2 \delta^2$?
 We follow the method described in \cite{peres} to estimate this bound.
 The same method
 is applied to estimate the $99\%$ CL upper limit on $\sin^2 2\vartheta_{LSND}$
 coming from Bugey (for $\gamma^2$) and CDHS and Super-K (for $\delta^2$),
 that is without using the SN1987A data.
\indent The results for the $(3+1)$ model with normal neutrino mass hierarchy
 and mixing given by Eq.~\ref{eqn:3+1} is shown in Fig.~\ref{fig:parspace}b.
 Also for this model, we find that existing constraints (the Bugey constraint
 on $\delta^2$, in this case) are stronger than the SN1987A one. \\
\indent Tab.~\ref{tab:summary} summarizes the SN1987A constraints
 obtained in this paper on the LSND allowed region, for the various
 neutrino mass and mixing models considered. 
\begin{table}[bt]
\begin{ruledtabular}
\begin{tabular}{lc}
Model & SN1987A constraint on\\
      & LSND region ($99\%$ CL) \\ \hline
Normal $(1+1)$ & partially excluded (Fig.~\ref{fig:parspace}a) \\
LSND-inverted $(1+1)$ & excluded \\
Normal $(2+1)$ & unconstrained \\
LSND-inverted $(2+1)$ & excluded \\
Normal $(2+2)$ &  partially excluded (Fig.~\ref{fig:parspace}a) \\
LSND-inverted $(2+2)$ & excluded \\
Normal $(3+1)$ & partially excluded (Fig.~\ref{fig:parspace}b) \\
LSND-inverted $(3+1)$ & excluded \\
\end{tabular}
\end{ruledtabular}
\caption{\label{tab:summary}Summary of the SN1987A constraints on the LSND
 allowed region, for the various models considered in this paper; see
 Fig.~\ref{fig:parspace} also.}
\end{table}
%
\section{\label{sec:four}CONCLUSIONS}
We have investigated the effect that 3- and 4-neutrino oscillation schemes
 would have in modifying the energy spectrum of supernova $\bar{\nu}_e$'s.
 Throughout the paper, we apply the adiabatic approximation for the
 antineutrino propagation in the supernova environment and neglect Earth
 matter effects. Moreover, we have used our results to test the 
 compatibility between the SN1987A data and the LSND evidence for
 $\bar{\nu}_{\mu}\rightarrow \bar{\nu}_e$ oscillations. \\
\indent We have provided specific relations for the permutation factor,
 which gives the admixture of a higher energy flux to the original
 $\bar{\nu}_e$ flux at production from $\bar{\nu}_{\mu},
 \bar{\nu}_{\tau}\rightarrow \bar{\nu}_e$ oscillations, for various neutrino
 mass and mixing models. The permutation factor may be measurable with good
 accuracy in future supernova experiments. \\    
\indent Based on SN1987A data only, which seem to indicate a small (if non-zero)
 value for the permutation factor, we are able to exclude all of the four
 models considered which would explain the LSND signal via a
 ``LSND-inverted'' neutrino mass hierarchy, as defined in the text.
 For the normal mass hierarchy schemes considered, SN1987A
 data do not provide any stronger constraints on the LSND allowed region
 for oscillations than those already obtained with reactor, accelerator
 and atmospheric neutrinos; additional experimental input
 is necessary to unambiguously discern the neutrino mass and mixing
 properties. Undoubtedly, the detection of supernova neutrinos by present
 or near-term experiments\cite{snexperiments} would prove very useful in this respect. 
\begin{acknowledgments}
 We thank Kevork Abazajian, Vernon Barger, John Beacom, Nicole Bell,
 Steve Brice, Klaus Eitel, Lam Hui, Hitoshi Murayama and Michael Shaevitz for valuable
 discussions and useful suggestions. This work was supported by NSF and
 by the Sloan Foundation.
\end{acknowledgments}
\appendix

\section{UPPER BOUNDS ON THE PERMUTATION FACTOR
 FROM SN1987A DATA}
In this appendix, we discuss the statistical methodology and the
 physics assumptions used to estimate
 the upper bound on the permutation factor $p$ quoted in the text,
 $p<0.22$ at $99\%$ CL. We use the same statistical methodology
 as in \cite{smirnov}, that is we use the
 Kholmogorov-Smirnov test on the joint Kam-IMB dataset to derive the
 upper bound. Most of the physics assumptions are identical to those in
 \cite{jegerlehner}. \\
The expected energy spectrum for the positrons, observed in the Kamiokande and IMB
 detectors via the reaction $\bar{\nu}_ep\rightarrow e^+n$, is:
\begin{widetext}
\begin{equation}
n_i(E_{det})=\frac{N_{p,i}}{4\pi D^2}\int_0^{\infty}\,dE_+ P_i(E_{det},E_+)\eta_{0,i}(E_+)
 \sigma_{\bar{\nu}_e p}(E_++Q)F_{\bar{\nu}_e}(E_++Q)
\end{equation}
\end{widetext}

\noindent where $i$ refers to either Kam or IMB, $N_{p,i}$ is the number of target protons in
 the detectors,
 $D$ the distance between the Large Magellanic Cloud and the Earth,
 $E_{det}$ ($E_+$) is the detected (true) positron energy, $Q\equiv m_n-m_p=1.29\mbox{MeV}\simeq
 E_{\bar{\nu}_e}-E_+$, $P_i(E_{det},E_+)$ and $\eta_{0,i}(E_+)$ the energy resolution functions
 and efficiency curves taken from \cite{jegerlehner}, $\sigma_{\bar{\nu}_e p}(E_++Q)\propto
 E_+^2$ the neutrino interaction cross-section taken from \cite{beacom} (neglecting
 nuclear recoil), and finally
 $F_{\bar{\nu}_e}(E_++Q)$ the neutrino flux at the detector taken from Eq.\ref{eqn:flux}.
 We assume ``unpinched'' Fermi-Dirac distributions for the fluxes $F^0_{\bar{\nu}_{\alpha}}$,
 $\alpha = e,\mu$, appearing in Eq.\ref{eqn:flux}:
\begin{equation}
F^0_{\bar{\nu}_{\alpha}}(E)\propto \frac{E^2}{\langle E_{\bar{\nu}_{\alpha}}\rangle
 T_{\alpha}^3(e^{E/T_{\alpha}}+1)}
\end{equation}
\noindent where $\langle E_{\bar{\nu}_{\alpha}}\rangle\simeq 3.15T_{\alpha}$ at the
 denominator ensures energy equipartition. \\
\indent The cumulative distribution function used for the Kholmogorov-Smirnov test is:
\begin{equation}
\mathcal{F}(E_{det})=\int_0^{E_{det}}\,dE(n_{Kam}(E)+n_{IMB}(E))
\end{equation}
\begin{figure}[bt]
\includegraphics{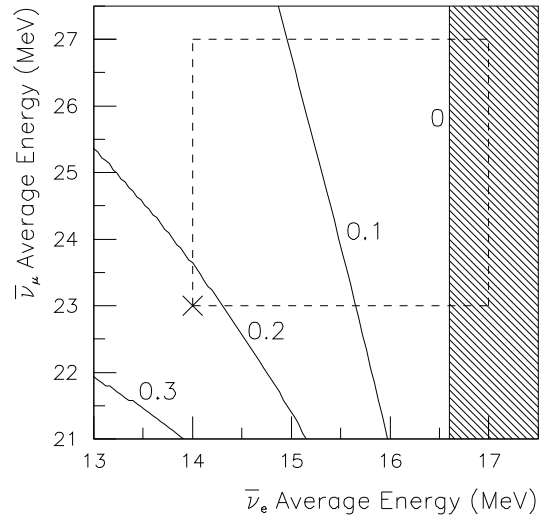}
\caption{\label{fig:sn1987bound} Solid lines: isocontours for the {\it upper bounds} on
 the permutation factor $p$ at $99\%$ CL obtained from SN1987A data, as a function
 of the $\bar{\nu}_e$ and $\bar{\nu}_{\mu}$ average energies predicted at production by
 supernova models;
 rectangle with dashed border: range of energies allowed by present
 models; cross: model chosen to derive the (conservative) upper bound on $p$ used
 in the text. The region $\langle E_{\bar{\nu}_e}\rangle>16.6$ MeV is excluded at $99\%$ CL
 for all values of $p$.}
\end{figure} 
\indent Fig.\ref{fig:sn1987bound} shows the upper bound on the permutation factor $p$
 obtained from SN1987A data, at $99\%$ CL, as a function of the average energies
 $\langle E_{\bar{\nu}_e}\rangle$, $\langle E_{\bar{\nu}_{\mu}}\rangle$. As expected,
 the bound becomes more stringent for supernova models in which the neutrino average
 energies are higher. SN1987A data are incompatible at $99\%$ CL with all supernova
 neutrino models predicting $\langle E_{\bar{\nu}_e}\rangle >16.6$ MeV, for
 all values of $p$ and $\langle E_{\bar{\nu}_{\mu}}\rangle$.
 We adopt a conservative approach, and quote as the upper bound on
 $p$ the largest value for supernova neutrino models in the range
 $14<\langle E_{\bar{\nu}_e}\rangle<17$ MeV, $23<\langle E_{\bar{\nu}_{\mu}}\rangle<27$ MeV,
 that is the one corresponding to $\langle E_{\bar{\nu}_e}\rangle=14$ MeV,
 $\langle E_{\bar{\nu}_{\mu}}\rangle=23$ MeV (cross in Fig.\ref{fig:sn1987bound}).
%

%


\begin{thebibliography}{99}
\bibitem{2+2nonadiab}Let us see what the effect of a possible violation of adiabaticity between the two highest
 energy eigenstates, $\bar{\nu}_{\mu}$ and $\bar{\nu}_{\tau}$ at high
 matter densities, in the case of a LSND-inverted (2+2) model (Fig.\ref{fig:respos}f) is
 on $p_{e\rightarrow e}$ and $p_{\mu\rightarrow e}+p_{\tau\rightarrow e}$. Violation of
 adiabaticity would result in a non-vanishing hopping probability $p_{hop}$ between
 these two local Hamiltonian eigenstates. Clearly, $p_{e\rightarrow e}$ is unaffected, and also:
$
 p_{\mu\rightarrow e}+p_{\tau\rightarrow e}=p_{hop}|U^{evol}_{e0}|^2+(1-p_{hop})|U^{evol}_{e1}|^2+
(1-p_{hop})|U^{evol}_{e0}|^2+p_{hop}|U^{evol}_{e1}|^2=|U^{evol}_{e0}|^2+|U^{evol}_{e1}|^2
$
 has the same value as in the adiabatic case.
\end{thebibliography}
\end{document}